\documentclass[manuscript]{acmart}
\renewcommand\footnotetextcopyrightpermission[1]{} % removes footnote with conference information in first column
\makeatletter
\renewcommand\@formatdoi[1]{\ignorespaces}
\makeatother

\AtBeginDocument{%
  \providecommand\BibTeX{{%
    \normalfont B\kern-0.5em{\scshape i\kern-0.25em b}\kern-0.8em\TeX}}}

\setcopyright{acmcopyright} % TOPUB fill this in
\copyrightyear{2020}
\acmYear{2020}
\acmDOI{10.1145/1122445.1122456}
\usepackage{color, colortbl}

\acmJournal{DTRAP} % TOPUB + numbers here
\acmVolume{0}
\acmNumber{0}
\acmArticle{0}
\acmMonth{0}

\begin{document}

%%
%% The "title" command has an optional parameter,
%% allowing the author to define a "short title" to be used in page headers.
\title{Vulnerability Forecasting: In theory and practice.}

%%
%% The "author" command and its associated commands are used to define
%% the authors and their affiliations.
%% Of note is the shared affiliation of the first two authors, and the
%% "authornote" and "authornotemark" commands
%% used to denote shared contribution to the research.
\author{\'{E}ireann Leverett}
\authornote{Both authors contributed equally to this research.}
\email{eireann.leverett@airbus.com}
\orcid{0000-0001-6586-7359}
\author{Matilda Rhode}
\authornotemark[1]
\email{matilda.rhode@airbus.com}
\author{Adam Wedgbury}
\email{adam.wedgbury@airbus.com}
\affiliation{%
  \institution{Airbus}
 \streetaddress{Quadrant House, Celtic Springs Business Park, Coedkernew, Duffryn}
 \city{Newport}
 \country{U.K.}
 \postcode{NP10 8FZ}
}

%%
%% By default, the full list of authors will be used in the page
%% headers. Often, this list is too long, and will overlap
%% other information printed in the page headers. This command allows
%% the author to define a more concise list
%% of authors' names for this purpose.
%\renewcommand{\shortauthors}{Leverett and Rhode}

%%
%% The abstract is a short summary of the work to be presented in the
%% article.
\begin{abstract}
    Why wait for zero-days when you could predict them in advance? It is possible to predict the volume of CVEs released in the NVD as much as a year in advance. This can be done within 3 percent of the actual value, and different predictive algorithms perform well at different lookahead values. It is also possible to estimate the proportions of that total volumn belonging to specific vendors, software, CVSS scores, or vulnerability types. Strategic patch management should become much easier, with this uncertainty reduction.
\end{abstract}

%%
%% The code below is generated by the tool at http://dl.acm.org/ccs.cfm.
%% Please copy and paste the code instead of the example below.
%%

\begin{CCSXML}
<ccs2012>
<concept>
<concept_id>10002978</concept_id>
<concept_desc>Security and privacy</concept_desc>
<concept_significance>500</concept_significance>
</concept>
</ccs2012>

\end{CCSXML}

%%
%% Keywords. The author(s) should pick words that accurately describe
%% the work being presented. Separate the keywords with commas.

%%
%% This command processes the author and affiliation and title
%% information and builds the first part of the formatted document.
\maketitle

\section{Motivation}

Most computer security vulnerability management problems are plagued by reactivity to the disclosure of a Common Vulnerability Enumeration (CVE) or the release of a patch. The cadence of patch release is necessarily correlated with the release of CVEs, while the patch installation is dictated by business risk patterns. The release of CVEs is commonly viewed as stochastic, unpredictable, and difficult to imagine before they emerge. Turning to expectation of exploits as posterior probabilistically from CVE disclosure: recent analysis has found that 80\% of exploits are written before the CVE is published, by 23 days on average \cite{chen}. This resonates with the authors' experience that a basic exploit often has to be written to convince the company of the need for a CVE in the first place. The only question is whether such exploits are released more publicly in frameworks such as metasploit, semi publicly on platforms such as exploitDB; or semi-privately in exploitkits.

A vision of the future of vulnerability and weakness management needs a solid foundation: the ability to forecast the volume of CVEs being produced within discrete time spans (ideally user specified). If those forecasts are reasonably accurate (plus or minus some confidence or prediction interval), then prediction of finer grain details will be accomplished more easily. Many other cyber risk management problems become significantly easier, because we have more time to plan ahead and thus become less reactive; we shift decision making to an earlier point in the timeline of vulnerability and risk management.

 The predicted volume of CVEs can be used both to inform patch strategies (priotitisation) but also resource requirement estimatation. Advanced planning of required resources for patching and vulnerability management would be valuable even if maximum optimisation of resources to risk reduction is not achievable. For example consider the severity, frequency, and types of vulnerabilities disclosed with forecasting in mind. It would be valuable to know if an exploit exists already, or the vulnerability was found in the wild already being exploited, regardless of our ability to minimise the risk at precisely the right moment and for the absolute minimum cost. Which is to say, a competent risk manager can manage the uncertainty if it is stated carefully in the forecasts.

A future application of CVE prediction will also be found in the conditional probabilities; such as  `If the next CVE in this software is a memory corruption, what probability is there that an exploit is developed within 3 months?' or  `If this software becomes deployed on 25\% of internet facing servers, what chance is there that an exploitable vulnerability will be found in it?'. To achieve any of that, forecasting and risk analysis of the CVE data, public disclosures, exploits, and operating systems, is necessary.

\section{Contributions}

The key contributions of this work are summarised below:
\begin{enumerate}
    \item A publicly available \textbf{leading indicator} that has not been used in previous work but drastically improves machine learning model accuracy (See ~\ref{indicator}). 
    \item Two statistical models are applied to this problem which have not been used in previous work and are shown to outperform currently leading machine learning and time-series forecasting approaches for long-term predictions (See ~\ref{two_models}). 
    \item A set of models for predicting the number of vulnerabilities in both the \textbf{short- and long-term}, forecasting within 3\% of the true values for 1 year lookahead predictions on average. Previous work has focused on short (1 to 3 months) or long term (1 or 2 years) time periods only, when patch managers are likely to want to analyse both next month's and next year's predictions (See ~\ref{framework}). 
    \item A flexible framework to predict the number of \textbf{total vulnerabilities or a specific subset}, using an appropriate model for the task. Previous work typically looks at the number of vulnerabilities for specific products only. They only cover well-represented products with a low CVE-instance variance. These are easier to predict (See ~\ref{cve_types}). 
    \item This work acknowledges that we may want to predict CVEs in products that have never had a CVE before. We uncover some challenges of predicting less commonly seen sub-types of vulnerabilities accordingly (See ~\ref{cve_types}).
\end{enumerate}

There are various different types of forecasting for security vulnerabilities. In this paper the focus is primarily on forecasting \emph{publicly disclosed vulnerabilities}, that might impact a user of any software or firmware. The aim is to translate that prediction and forecasting capability to optimise minimum resources for maximum risk reduction in vulnerability management. However, we take a broader view in the literature below to overview other aspects of the research because it also offers insights. So what has been done to date in this respect?

\section{Background and Related Work} % renamed because it has background + lit review

As a lightweight taxonomy, there are three broad streams of vulnerability forecasting research. 

\begin{enumerate}
    \item \textbf{Existing network vulnerabilities} Given previous network or single host vulnerability scans, forecast how many previously known CVEs will be next found, and the density of vulnerabilities in different targets, or more simply: the rate of change between host or network scans.
    \item \textbf{Future software vulnerabilities} Given source code or a compiled binary, forecast how many vulnerabilities there might be, where they are, or how much effort it might take to find them.
    \item \textbf{Future vulnerabilities: all types} Given data on historical Common Vulnerability Enumerations (CVEs), use machine learning to estimate the volume of published vulnerabilities in a given time frame, and/or resource requirements to patch them. Given further information such as Common Vulnerability Scoring System (CVSS) or knowing an exploit already exists, estimate difficulty to patch and/or risk reduction from having done so.
\end{enumerate}

In the literature review that follows the papers are grouped into these three broad categories, skipping briefly beside the first two streams, before diving into the rushing rapids of the third.

\subsection{Stream One: Forecasting the Vulnerability of Specific Computers or Networks}

This branch of academic literature concerns itself with the prediction of the vulnerability of a given machine or network. Sometimes it is even more specific, such as the presence of a particular Common Vulnerability Enumeration (CVE) on an Internet Protocol (IP) address or range of addresses e.g. \cite{venter2004vulnerability}. Or, closely related to the CVE forecasting work in this paper, estimating the number of unpatched vulnerabilities in an organization based on the products in use and Bayesian inference \cite{chatzipoulidis2015information} or fuzzy sets \cite{ALEKSIC2014214}. This seems more applicable to red teams than blue teams, though blue teams could use it as a metric to optimise patching strategies. For example, if you wanted to patch machines non-homogenously across a network, to make it hard to predict which machines were vulnerable to different exploits. This may have stronger application where scanning is likely to miss results due to computers joining and leaving the network (University WiFi) or being down for maintenance (OT Networks). In short, where the network's graph of connections could be considered a dynamic in temporal sense, this type of prediction may be very useful.

\subsection{Stream Two: Pre-disclosure Vulnerability Forecasting}

There is plenty of literature on predicting the number of undiscovered/undisclosed vulnerabilities in software. Some of that literature focuses simply on the numbers \cite{vander1993assessing, petersson2004capture, scott2008capture}, and other papers on how much effort it takes to find them and what improvement may or may not be achieved \cite{brady1999murphy, rescorla2005finding, ozment2006milk}. Some even make forays and efforts into the value of predictability and forecasting \cite{alhazmi2006prediction}. The primary aim and application of this type of research is the prevention of vulnerabilities during the software development lifecycle. In other words, preventing the vulnerabilities from reaching a production system and getting a CVE at all.

Another distinctive sub-field of this research advocates becoming active; one can deliberately inject vulnerabilities to be able to figure out the rate at which your security testers find them \cite{mci/Schuckert2016}. Alternatively, one could use a security code review team to find vulnerabilities, then not fix them, and see if the QA team finds them again to produce an estimation of how many have not been observed (Lincoln Index) \cite{geer2015good}. Such an index allows you to estimate the vulnerabilities remaining in a static codebase, which may or may not have value depending on the velocity of your development team; if your development team is changing the code too quickly, your fixes may not have much value as they inject more vulnerabilities at a higher speed.

%% commented out to save space:
%This research is much more mature than the other two streams, and some other notable examples are: TODO

%In summary this stream is directed towards predicting pre-release vulnerabilities in specific code bases.

\subsection{Stream Three: Post-Disclosure Vulnerability and Exploit Forecasting}

This is the stream of greatest relevance for our own work within this paper, and explores the predictive power of estimating the number of vulnerabilities that are publicly disclosed as CVEs. 

The primary application of such research would be in managing patching teams, tools, and networks, exploring patch strategies, and thinking about third party software interactions. 

Yasasin et al present an excellent systemization of knowledge paper, that also demonstrates some promising  and practical empirical results~\cite{yasasin2020forecasting}. The paper surveys the literature much more broadly than we do here, and provides many other points of departure for the interested reader. The authors compare multiple time series forecasting models to predict the number of vulnerabilities for specific vendors. In particular if the model under construction is meant to be vendor or product specific one (I.E. all Microsoft product disclosures, or Mozilla:Firefox disclosures), for which there may be many time periods in which there are zero disclosed vulnerabilities. The authors caution against the use of relative metrics such as percentages. The authors detail how to test the predictive power of various time series models and report that Croston's and ARIMA forecasting models tend to perform best but do not select an overarching methodology for predicting the number of vulnerabilities attributed to \textit{any} product. Further the work is limited to 1, 2 and 3 month forecasts "Since forecasting research has shown that long-term forecasts are generally limited in their predictive accuracy", whereas here we seek to predict up to one year in advance. Our work builds on this paper by using a more robust testing methodology where strict date separation is enforced between the training and test set to better-mimic real-world use (more details in Subsection~\ref{evalutation_methodology}).

Another inspiring paper revisits predicting product specific disclosures for Operating Systems~\cite{pokhrel2017cybersecurity}. This one details both ARIMA and an approach using Artificial Neural Networks (ANNs). The authors predict numbers of CVEs for 12 single months, sum them, and compare to the yearly total for the test set, but this is not the same as predicting the next year's total CVEs since the model has information about what has happened so far in a year (e.g in July) to predict the total number appearing between January and December. These increasing forecast projections are decreasingly accurate, but might still be accurate enough for resource planning. Here the authors observe strict date splitting, only testing on data from 2016 and training on data up to the end of 2015. The confidence intervals are defined for the predictions, which also assist the astute risk manager. However, we note that prediction intervals would have been a better choice to represent uncertainty boundaries, than confidence intervals in this respect as the former capture both the uncertainty around the mean prediction \textit{and} around the variance (thus the prediction interval is always wider than the confidence interval). 

Last \cite{last2016forecasting} predicts broader categories: all vulnerabilities, Browser CVEs, Operating System CVEs, and Video Codec CVEs. This approach uses root mean squared percentage error (RMSPE) as a validation metric (which the deep reader remembers was discouraged by \cite{yasasin2020forecasting} in the cases of zero-inflated time series); and notes the occasional limited success of using software release dates alongside other data to aid in the forecasting accuracy. This work predicts 12, 18 and 24 month windows, using strict date separation, with a testing window from 2012 to 2015, achieving a median 15\% RMSPE for global vulnerability predictions over 24 months, the work presented herein is able to significantly improve this metric for long-term global predictions. Last's work reveals an increasing accuracy as the forecast window increases from 12 to 24 months, this is a result we have found to extend to short term predictions: the total number of vulnerabilities next year is easier to predict than next month's. However, this approach predicts the cumulative total CVEs for all time, such that the percentage error metrics cannot be considered consistent over time, furthermore the cumulative totals i.e. how many CVEs there have been since 2001, are unlikely to be useful to patch managers. 

At the more mathematical end of the academic spectrum, we find a paper which provides methods to optimise response to demand volumes \cite{4721516}. This provides a solid foundation for optimising our response to the numbers of vulnerabilities a potential model may predict but, more discussion is needed about the variability of response to differing numbers of patches. Useful as this paper is, it is not specific to our domain of publicly disclosed vulnerabilities, though it does detail an interesting method for making unsupervised hybrid models by scaling horizontally with more processing power. In this research, we have not found processing power to be a limiting factor.

One paper notes that vulnerabilities follow market share~\cite{4530404}, while another warns us to `mind the denominator'~\cite{doi:10.1080/23738871.2018.1472288}. For our purposes that might mean using the full Common Platform Enumeration (CPE) dictionary to look at probabilities for vulnerabilities, instead of only vulnerability databases. Indeed we discuss later in the paper the challenge of predicting CVEs for products that have never appeared in the CVE List before, and that is where using the CPE might help in some cases. 

It is important to scour amongst hacker literature as well as academic papers. Many academics hide their light under a bushel and present very practical results at hacker conferences such as BlackHat or Chaos Communication Congress. Allodi, for example, has captured a very important piece of information security risk reduction measurement in this presentation of the perils of CVSS as a risk quantification\footnote{\url{https://media.blackhat.com/us-13/US-13-Allodi-HOW-CVSS-is-DOSsing-Your-Patching-Policy-Slides.pdf}}. The gist of it is that most vulnerabilities don't have exploits, and don't get exploited. Improving the \textit{specificity} (false positive for exploited vulnerabilities) of our work would save us a lot of patching effort. Allodi has also published a related paper on the distribution of exploitation per cve\cite{HeavyExploitation}.

An intellectual offspring of Allodi's work was performed by Cyentia Institute and Kenna Security\footnote{\url{https://i.blackhat.com/USA-19/Thursday/us-19-Roytman-Predictive-Vulnerability-Scoring-System-wp.pdf}}. Here they look at a wider dataset of exploited machines in IDS traffic, trying to learn from what gets exploited and extrapolating towards better patching. That work has now become a special interest group at FIRST.org for Exploit Prediction and Scoring Systems (EPSS). 

Previous work into vulnerability forecasting has demonstrated that it is possible to predict the numbers of all expected CVEs or subsets of CVE categories for a few months or a couple of years but has failed to develop a framework that is capable of spanning both. In this paper we present a model to predict both short- and long-term numbers of vulnerabilities using publicly available data from MITRE that has not previously been written about for this purpose.

This stream also has applications to supply chain management and passive vulnerability scanning as well. It may not be intuitive that forecasting is useful to risk management even when you have \textit{no intention} of patching as a risk treatment. To illustrate that it is still useful, consider that CVE forecasting could answer questions of the form: A submarine is launched with vendor X's product; knowing it can not be patched for a year due to cost/battery/bandwidth reasons, how many vulnerabilities can be expected to emerge in Vendor X's product within that year of service? Another way of looking at it might be: Estimate how often I need maintenance downtime to patch Vendor X's products above CVSS 7 this year. With this in mind, what data can we use for prediction and forecasting, to move towards answering such questions?

\section{Data Used}

As with previous work in vulnerability forecasting \cite{last2016forecasting, pokhrel2017cybersecurity, yasasin2020forecasting} this research uses the US-government's National Vulnerabilities Database (NVD) with a goal to predict the number of vulnerabilities (matching certain criteria) that will be published to this database within a given future time span (lookahead window). The NVD data itself can be used as a signal for this prediction but is more effective when augmented by additional data from MITRE; the organisation that maintains and publishes the CVE-list, which feeds the NVD. 

\subsection{NVD}

The NVD maintains detailed information about CVEs including but not limited to:

\begin{enumerate}
    \item date published
    \item the organisation reporting the CVE (known as a CVE Numbering Authority or CNA)
    \item a text description
    \item Strings indicating specific entities impacted by the CVE (Known as Common Platform Enumerations or CPE strings)
\end{enumerate}

\begin{figure*}[h!]
    \includegraphics[width=0.9\textwidth]{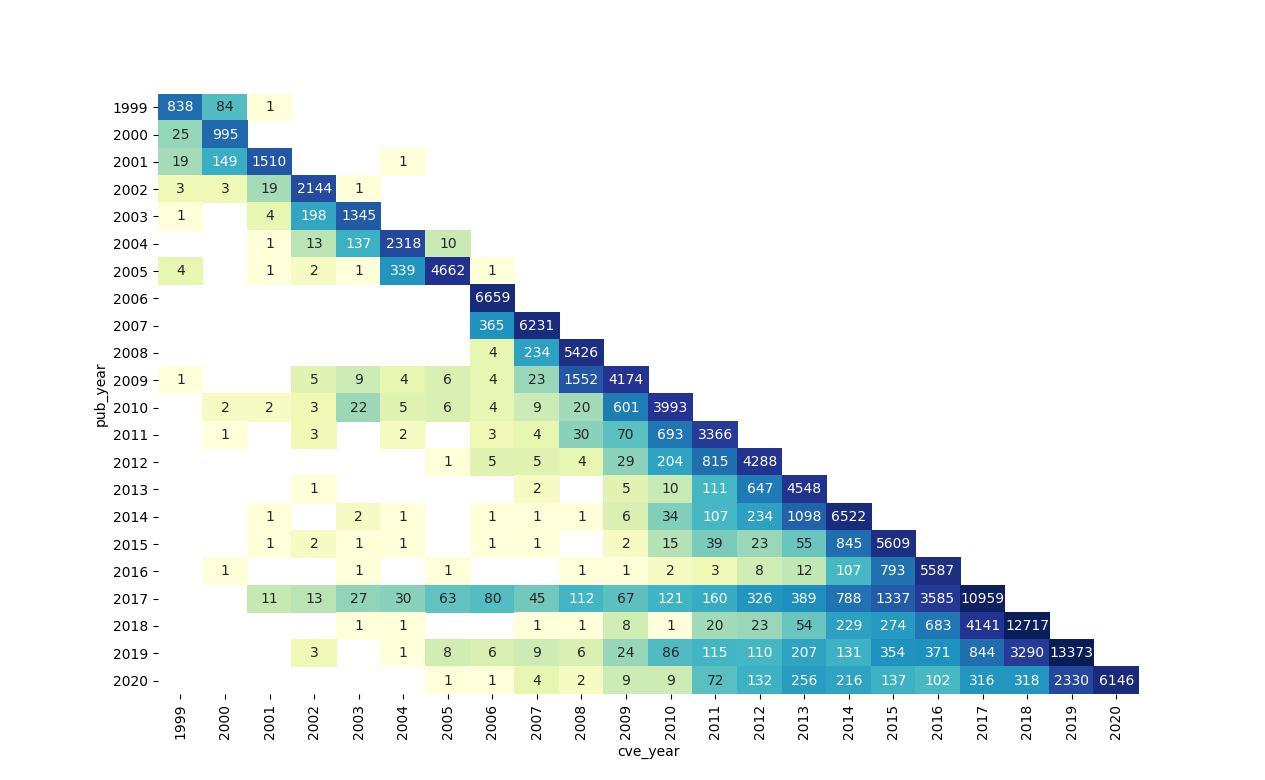}
    \caption{Heatmap of CVE Year against Publication Year over the historical record.}
    \label{fig:Heatmap}
\end{figure*}

CVEs are identified by a CVE-ID string in the format: CVE-YEAR-XXXX. During this research it became evident that the YEAR in the CVE-ID did not match the publication year in all instances. This is because the CVE-ID reflects the year in which the vulnerability was \textit{assigned} rather than published. Figure~\ref{fig:Heatmap} illustrates the disparity between publication year and CVE-ID year, some early CVEs were added retrospectively even assigned a CVE year later than they are published but this is an anomalous attribute of the early entries in the database. Every CVE is assigned prior to being published, sometimes years in advance, sometimes as little as an hour.  The XXXX part of the ID corresponds to a serial number thus allowing us to infer that on publication of CVE-2020-0100 there are unpublished CVE IDs reserved for CVE-2020-0001 to CVE-2020-0099 even if they have not been published\footnote{In practice these can be more digits such as CVE-2020-XXXXX or CVE-2020-XXXXXX}. Thus we are able to find that $N$ CVEs are in the pipeline, which improves the accuracy of predictive models. This latent data is the foundation of one of the models presented in this paper in \ref{matrix}. In short, CVEs are ordered on arrival, but released stochastically according to progress internal to MITRE. That ordering allowed us an initial insight into how long they take to process, later supplanted by the CVE list we discuss below.

\subsection{MITRE}

\begin{table}[!h]
    \centering
    \begin{tabular}{l|r}
        mean & 162 days 11:50:28 \\
        standard deviation & 390 days 06:14:18\\
        min & 0 days 01:02:00 \\
        1st quartile & 1 days 05:17:00 \\
        median & 51 days 15:57:00 \\
        3rd quartile & 160 days 20:29:00 \\
        max & 6263 days 15:15:00 \\
    \end{tabular}
    \caption{Summary of lags between database entry and publication of CVEs}
    \label{tab:lag_summaries}
\end{table}

Every CVE-ID is assigned by MITRE or one of their approved CNAs\footnote{144 organisations as of 5th November 2020: \url{https://cve.mitre.org/cve/cna.html}}. The specific assigned dates are published by MITRE \footnote{Available to download: \url{https://cve.mitre.org/data/downloads/index.html}}, which gives much better granularity of the assigned date than using the CVE-IDs. MITRE provides the assigned date whilst the NVD catalogues the date published. Research from Chen (PaoloAlto Networks) \cite{chen} asserts that 80\% of CVEs have a publicly available exploit on average 23 days before the CVE is published. Combining NVD and MITRE data reveals that the median lag between assigned date and publication date is 51 days (see Table~\ref{tab:lag_summaries}). These statistics underscore the possibility and usefulness of being able to predict the publication of CVEs at least one month in advance.

\subsection{Challenges}

Examining the raw data highlights some of the challenges in predicting CVE numbers. Four key challenges are briefly described below:

\begin{figure*}[!h]
    \centering
    \includegraphics[width=0.9\textwidth]{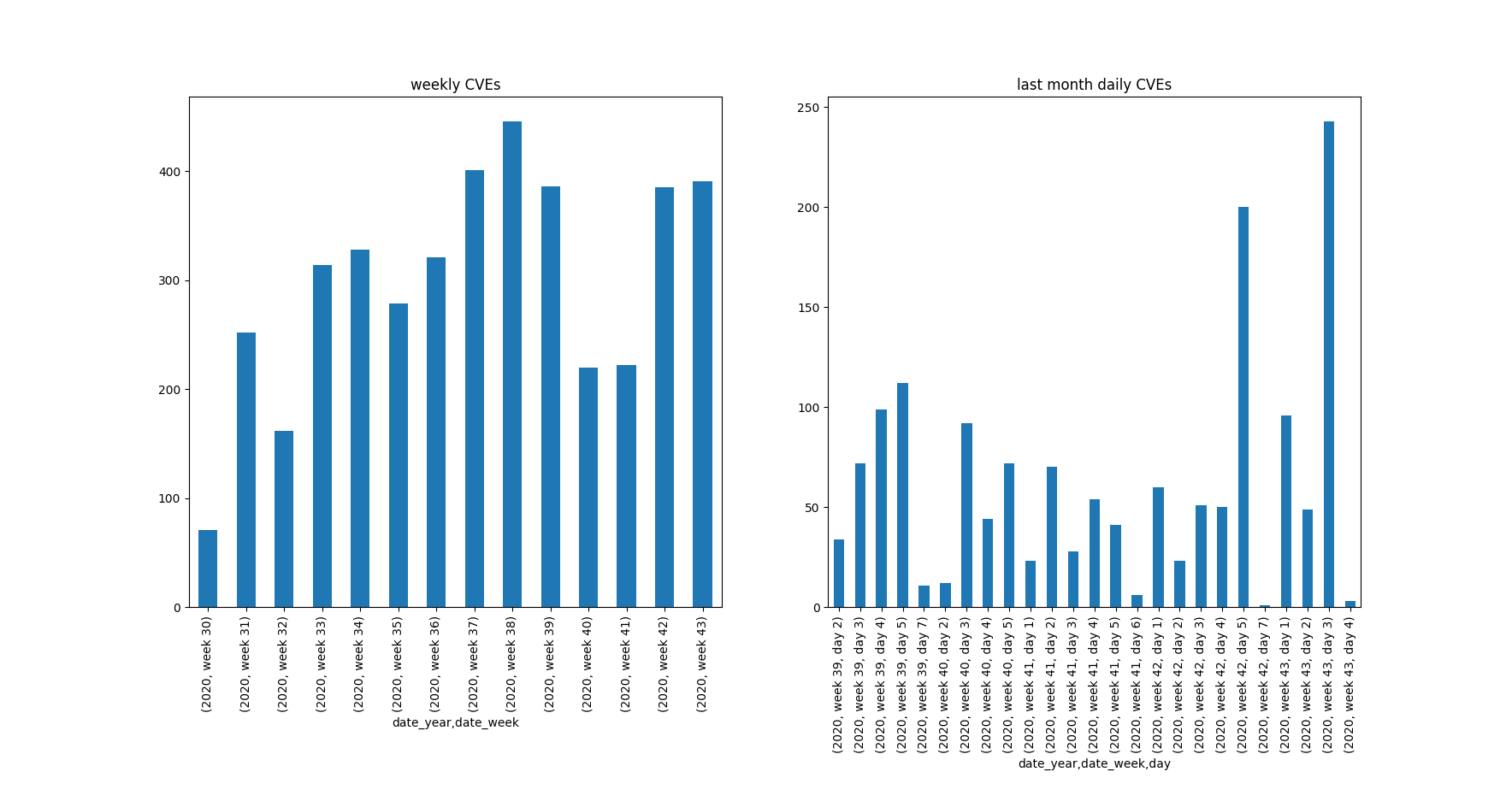}
    \caption{Number of CVEs published in weeks 42-46 of 2020 per week (left) and per day (right)}
    \label{fig:short_term_variance}
\end{figure*}

\textbf{Short-term variance} Due to the idiosyncrasitic policies and practices of both CNAs and MITRE, some of which are common to all non-automated processes, sometimes an abormally large number of CVEs are published in a single day. Figure~\ref{fig:short_term_variance}) shows the number of CVEs published on a daily and weekly basis. Some days there are not any CVEs published e.g. there are few Saturdays, (day 6s) in the graph. One day there are 243 published. This high variance is less evident in the aggregated week publications and there is less variance still in monthly and yearly publication numbers, illustrated by Figure ~\ref{fig:normed_variance} . This presents a challenge for predicting short-term numbers of all CVEs. 

\textbf{New products} Some products may see a sudden, unprecedented increase in the number of CVEs. This may be because the product is new, because it's market share has increased \cite{4530404} or due to some other factor that is exogenous to NVD or MITRE data. It is challenging for the model to predict something entirely new which has either not previously existed or has been at zero for a long time. This is a challenge for long-term CVE predictions.

\begin{figure}
    \centering
    \includegraphics[width=0.5\textwidth]{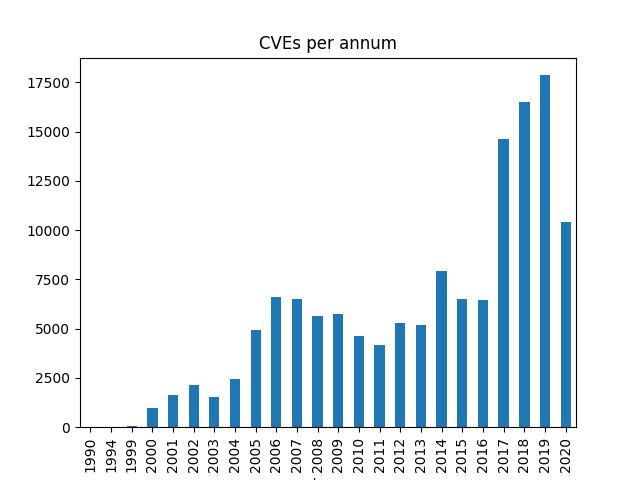}
    \caption{Annual total CVEs published 2002 to 2020}
    \label{fig:annual_counts}
\end{figure}

\textbf{Structural changes in CVE publication patterns} Though the variance is lower for annual total CVE counts, there are clear step-changes in the numbers of CVEs being published (see Figure~\ref{fig:annual_counts}) either side of 2005 and of 2017 and to a lesser degree in 2014. This is a challenge for models using historic data to predict new vulnerability totals, one which is explored using data filtering in the experiments reported in this paper.

\textbf{Assigned date is not a sufficient leading indicator} for long term prediction. Whilst the assigned date will indicate how many CVEs are in the pipeline, this data is not sufficient to predict the number of vulnerabilities that will be published within a given forecast window (lookahead). This is because (1) some CVEs are rejected by MITRE (2) some CVE-IDs are reserved by CNAs speculatively but never populated (3) even if we know how many are in the pipeline, we do not know how many more will be assigned \textit{and} published within the forecast window. 

\subsection{Evaluation} \label{evalutation_methodology}

This section presents the two evaluation methodologies used in this paper: retrocasting and forecasting. Retrocasting is used for evaluating model accuracy, forecasting is for resource management and planning, i.e. real-world use. 

Some previous work e.g \cite{yasasin2020forecasting} that has not separated training and test sets by date may have inadvertently leaked data about yearly structural changes or the existence of new products in the NVD database, in reality a model would have no warning of these changes. For retrocasting, as strictly future time window is used: September 2018 to September 2020. To calculate the prediction interval we use a validation set from January 2018 to September 2018. Training data precedes 2018. The prediction intervals were used to provide \(\pm\) bounds around the forecasted values. 

\textbf{Retrocasting} iteratively pretends we are somewhere in the past trying to predict the number of CVEs in the future, this approach allows us to approximate the performance of our model when it is deployed. Retrocasting predicts the number of CVEs seen in a fixed period of time (e.g. one month) for multiple ``end" dates (end of the lookahead window) in the test set, here the end dates are staggered at 10 day intervals between Sept. 2018 and Sept. 2020. Retrocasting fixes the lookahead (the window of time we are trying to predict) but changes the end date and use the average accuracy metrics as an indicator of future performance. Empirically we verified that it is possible to be more accurate on some longer time windows, than on some shorter ones. Strange, but true.

\textbf{Forecasting} by contrast fixes the date from which we are trying to predict and changes the lookahead to give estimates for different time periods into the future. The purpose of forecasting is that for use in resource management, patch maintenance teams may want a short-to-long-term view of the number of vulnerabilities being seen. 

Previous work has indicated that a range of machine learning and statistical paradigms, some capable of time series analysis are superior to others in their forecasting capabilities. For comparison with prior work and to ensure that the most accurate model is being used for each forecast context (lookahead and CVE sub-group) a number of evaluation metrics are collected to determine the best model. 

The metrics used in this paper are outlined in the following table for \(N={i=0,...N}\) prediction windows where \(y_i\) is the true number of CVEs published in a single window and \(\hat{y_i}\) is the corresponding predicted number:

\begin{itemize}
    \item Mean absolute error (MAE): \(\frac{1}{N}\sum_{i=0}^{N}(|y_{i} - \hat{y}_{i}|)\) gives average absolute difference between true values and predicted values
    \item Mean squared error (MSE): \(\frac{1}{N}\sum_{i=0}^{N}(y_{i} - \hat{y}_{i})^{2}\) exponentially penalises large differences
    \item Mean logarithmic squared error (MSLogE): \(\frac{1}{N}\sum_{i=0}^{N}(log(y_{i}+1) - log(\hat{y}_{i}+1))^{2}\) does not penalise large differences disproportionately if the values themselves are large (e.g. 1000 vs 1100 is not as bad as 0 vs 100) and penalises underestimates more than overestimates.
    \item Prediction interval: Let \(\sigma\) be an unbiased estimate of of the standard deviation calculated by \(\sqrt{\frac{\sum{(Y - \hat{Y})^2}}{N-2}}\), and \(z\) a confidence interval z-score  e.g. 1.96 for 95\% confidence interval, then the prediction interval is \(\hat{Y}\pm z*\sigma\) for a validation set 
    \item[]{\textbf{Custom metrics}}
    \item Percentage of predictions more than 10\% from true values: \(\frac{1}{N}\sum_{i=0}^{N}(\frac{|y_{i} - \hat{y}_{i}|}{y_i}  > 0.1)\)
    \item Percentage of predictions more than 5\% from true values: \(\frac{1}{N}\sum_{i=0}^{N}(\frac{|y_{i} - \hat{y}_{i}|}{y_i} > 0.05)\)
    \item Percentage of predictions more than 5\% below true value \textit{or} 20\% above true value: \(\frac{1}{N}\sum_{i=0}^{N}((\frac{\hat{y}_{i} - y_{i}}{y_i} > 0.2) * (\frac{y_{i} - \hat{y}_{i}}{y_i} > 0.05)))\)
\end{itemize}

The first three provide benchmarking against existing approaches, and even some verification of their results. Of the mean metrics, this paper prefers MSLogE as it scales the difference between true and predictive values according to the magnitude of the true value e.g. observed 1000 vs predicted 1100 is less bad than observed 0 vs predicted 100, MSE and MAE give the same error for both. MSlogE also penalises underpredictions more harshly that overpredictions, in this use case, we would rather overestimate the resources required for vulnerability management for next year than underestimate them. The custom metrics appeal to the notion that in practice patch managers may care more about how often the model is right rather than the numeric accuracy over all time. After all, they've lived with uncertainty for nearly two decades already.

\section{Predictive models}\label{two_models}

For comparison against previous work we use the following models. \textbf{Non-time-series statistical models:} Single back-cast: take the previous year's (month, week, etc.) total to predict this year (month, week, etc.); rolling mean: Take a mean over the last few years (months, weeks, etc.). \textbf{Non-time-series machine learning models:} Logistic regression,  Bayes Ridge, Random Forest, Multi-layer perceptron (feed-forward neural network. \textbf{Time-series statistical models:} ARIMA, Croston's. \textbf{Time-series machine learning models:} Long-Short-Term Memory (LSTM) recurrent neural network. 

The line between a statistical and machine learning model is not sharply defined. Here, we distinguish machine learning and non-machine learning models as those models capable of learning from multiple input features and those which only take the historical data of the target values as inputs. 

We further present two models \textbf{Minimum Variance Unbiased Estimator}: predicts the total number of CVEs based on the highest ID number seen and the number published; \textbf{Little's Law with Lookback}: Uses elements of Little's law and predicts the entry rate into the MITRE database, scaling this by the ratio of CVEs published within a given window following entry. The next two sections outline these models in full detail.

\subsection{A Yearly MVUE} \label{indicator}

A well known Minimum Variance Unbiased Estimator exists for serial number prediction problems. It is: 

\begin{math}
    N \approxeq (M/M*Obs)-1
\end{math}

Where \(N\) in our case would be the number of CVEs with a single year as string name (CVE-2020-XXXX for example). \(M\) is the maximum CVE published, and \(Obs\) is the number of CVEs published (observed) that conform to the string CVE-2020-XXXX. Obviously this gives us another challenge, which is to maintain the numbers for all previous years, and see if they have all been published yet. After all, it is evident that some vulnerabilities appear 17 years later~\cite{cve20022444}.  

This indicator was in important stepping stone for our efforts, because it enabled us to evaluate the proportion of CVEs that get published within the same year and the length of time it takes to publish all CVEs assigned within a year. Specifically, note that it represents all the CVEs assigned in that year (actually we believe they start assigning them from Sep the previous year), and that some of those will still be published years afterward. So in this sense, it is not a good forecast model, but serves us better as an indicator.

It gave crucial insights which led to the next approach, trying to manage the state of predicted and observed CVE totals in a simpler way. We expect that many vulnerability management organisations already knew this trick, but for some reason we can't find reference to it in academic literature, and so we publish it here.

\subsection{A Matrix Based Approach} \label{matrix}

The best way to explain this methodology is with some tables to represent a large stored matrix. The goal is to capture a few important transitions of the matrix based approach that updates probability vectors to reflect new information, and makes uncertain events certain after they occur.

Imagine it is near the end of 2019, and a vulnerability comes out marked CVE-2019-0002. This vulnerability came out in 2019, so we put a 1.0 in this column to represent this certainty (Table~\ref{tab:intro_mvue_matrix}). We can also infer the existence of CVE-2019-0001 as described in the section above, and write it into a row as well. However, we don't know when this one will come out, but we can use the historical record to say 94 per cent come out within the same year, and 5 percent come out the next year, adding in the probabilities created from the historical record across as many years lookback as we think is relevant.

\begin{table}[]
    \centering
    \begin{tabular}{||c|c|c|c|c||}
       \hline 
       \hline 
       CVE Name & 2019 & 2020 & 2021 & 2022 \\
       \hline 
       CVE-2019-0001 & 0.94 & 0.054 & 0.0032 & 0.0028 \\
       CVE-2019-0002 & 1.0 & 0.0 & 0.0 & 0.0 \\
       \hline 
       \hline 
    \end{tabular}
    \caption{CVE-2019-0002 is published, and we can infer the existence of CVE-2019-0001. }
    \label{tab:intro_mvue_matrix}
\end{table}

Now imagine that 2019 has come and gone. We know that CVE-2019-0001 has not been published in 2019. So we need to update our probabilities to reflect this new information. In this example, we have used a made up set of probabilities = [0.94, 0.054, 0.003, 0.002], which for the sake of a simpler discussion let's assume come from a historical record of 1000 vulnerabilities over all time\footnote{The real vectors can be seventeen years/items long and run to many more decimal places.In practice only the first few years make up enough proportion to be relevant to accuracy.}. First we zero out the remainder of the row, of the CVE probabilities (see Table~\ref{tab:initial}).

\begin{table}[]
    \centering
    \begin{tabular}{||c|c|c|c|c||}
       \hline 
       \hline 
       CVE Name & 2019 & 2020 & 2021 & 2022 \\
       \hline 
       CVE-2019-0001 & 0.0 & 0.0 & 0.0 & 0.0 \\
       CVE-2019-0002 & 1.0 & 0.0 & 0.0 & 0.0 \\
       \hline 
       \hline 
    \end{tabular}
    \caption{CVE-2019-0001 is not published in 2019.}
    \label{tab:initial}
\end{table}

Then we must calculate the new probability by re-scaling the vector across the remaining relevant probabilities. Let's say 940 of our vulnerabilities are published the same year as they are assigned, but now we must eliminate them from our study. That leaves a historical record that is 54 published 1 year later, 3 published 2 years later, and 2 published 3 years later [54, 3, 2]. We sum the remaining list and get 59 vulnerabilities\footnote{The numerate amongst you will realise 940+59 does not equal 1000, and that can reflect the number of vulnerabilities that go unpublished, disputed, or rejected. Of course it is also possible to use only the vulnerabilities that get published and provide a another term to estimate the percentage that will get published, but in practice it makes little difference how you approach that. As long as the method is consistent, it will produce similar results.}. So our updated probability becomes [54/59,3/59,2/59] or [0.91,0.05,0.03] (see Table~\ref{tab:rescaling}).

\begin{table}[]
    \centering
    \begin{tabular}{||c|c|c|c|c||}
       \hline 
       \hline 
       CVE Name & 2019 & 2020 & 2021 & 2022 \\
       \hline 
       CVE-2019-0001 & 0.0 & 0.91 & 0.05 & 0.03 \\
       CVE-2019-0002 & 1.0 & 0.0 & 0.0 & 0.0 \\
       \hline 
       \hline 
    \end{tabular}
    \caption{CVE-2019-0001 is not published in 2019.}
    \label{tab:rescaling}
\end{table}

This method of rescaling probabilities works well, and is rather efficient. You just update the whole table once a year. Now let's examine another important scenario, new vulnerabilities being added for 2020. Let's imagine CVE-2020-0002 and CVE-2020-0004 are published, remember this means we can infer the existence of CVE-2020-0001 and CVE-2020-0003, to which we assign a probabilistic publication (see Table~\ref{tab:prediction}).

\definecolor{LightCyan}{rgb}{0.90,1,1}
\newcolumntype{g}{>{\columncolor{LightCyan}}c}
\begin{table}[]
    \centering
    \begin{tabular}{||c|c|g|c|c||}
       \hline 
       \hline 
       CVE Name & 2019 & 2020 & 2021 & 2022 \\
       \hline 
       CVE-2019-0001 & 0.0 & 0.91 & 0.05 & 0.03 \\
       CVE-2019-0002 & 1.0 & 0.0 & 0.0 & 0.0 \\
       CVE-2020-0001 & 0.0 & 0.94 & 0.05 & 0.03 \\
       CVE-2020-0002 & 0.0 & 1.0 & 0.0 & 0.0 \\
       CVE-2020-0003 & 0.0 & 0.94 & 0.05 & 0.03 \\
       CVE-2020-0004 & 0.0 & 1.0 & 0.0 & 0.0 \\
       \hline 
       \hline 
    \end{tabular}
    \caption{CVE-2020-0002 and CVE-2020-0004 are published in early 2020.}
    \label{tab:prediction}
\end{table}

Now, let's predict the number of Vulnerabilities that will be published. Sum the column for 2020 (highlighted in cyan in Table~\ref{tab:prediction}), to predict for 2020. That reflects accurately how many were published, but also how many are expected. In this case the answer would be 4.76, rounded up to 5.

How do we calculate those probabilities from this matrix? Let's publish one more vulnerability before we do that, to illustrate.

Now suppose CVE-2019-0001 finally gets published, and we record this by zeroing out the probabilities and recording the certainty with a 1.0 in the 2020 column, next to the CVE-2019-0001 row (see Table~\ref{tab:probabilities}).

\begin{table}[]
    \centering
    \begin{tabular}{||c|c|c|c|c||}
       \hline 
       \hline 
       CVE Name & 2019 & 2020 & 2021 & 2022 \\
       \hline 
       \rowcolor{LightCyan}
       CVE-2019-0001 & 0.0 & 1.0 & 0.00 & 0.0 \\
       \rowcolor{LightCyan}
       CVE-2019-0002 & 1.0 & 0.0 & 0.0 & 0.0 \\
       CVE-2020-0001 & 0.0 & 0.94 & 0.05 & 0.03 \\
       \rowcolor{LightCyan}
       CVE-2020-0002 & 0.0 & 1.0 & 0.0 & 0.0 \\
       CVE-2020-0003 & 0.0 & 0.94 & 0.05 & 0.03 \\
       \rowcolor{LightCyan}
       CVE-2020-0004 & 0.0 & 1.0 & 0.0 & 0.0 \\
       \hline 
       \hline 
    \end{tabular}
    \caption{CVE-2019-0001 is published in mid 2020, and probabilities are calculated.}
    \label{tab:probabilities}
\end{table}

\begin{figure}
    \centering
    \includegraphics[width=0.4\textwidth]{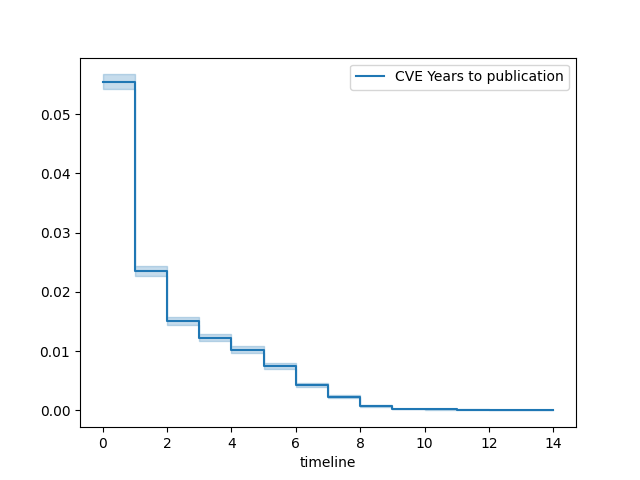}
    \caption{Kaplan-Meier Years.}
    \label{fig:KMYears}
\end{figure}

To recalculate the probabilities now is relatively easy. We make a filter on the table to only include rows that have 1.0 within them. This ensures we don't tabulate with vulnerabilities not yet published. For each column you simply subtract the cve-string-year from the column name, which gives you the year lag after publication. This is used as an index to build a vector of year lags\footnote{Doing so for months is only slightly more complicated, involving many more columns, but the method is the same.}. So for this table alone that would be: [3,1,0,0]\footnote{Once again the quantitative reader will already have figured out we needed to include our assumed history and the vector is actually [943,55,3,2,0]}. Now simply sum the list and divide each element by the sum to make a proportional list: [3/4,1/4,0/4,0/4]. This approximates Fig.~\ref{fig:KMYears} very accurately with minimum effort and updates.

Another way of viewing that same story is across the entire historic record, see Fig~\ref{fig:Heatmap}. Such a view allows us to see very interesting historical context such as the 2017 push to publish old vulnerabilities. 

\begin{figure}
    \centering
    \includegraphics[width=0.4\textwidth]{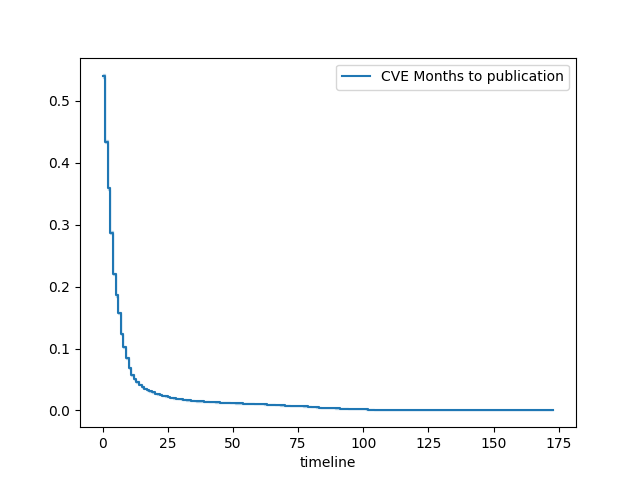}
    \caption{Kaplan-Meier Months.}
    \label{fig:KMMonths}
\end{figure}

\begin{figure*}
    \centering
    \includegraphics[width=\textwidth]{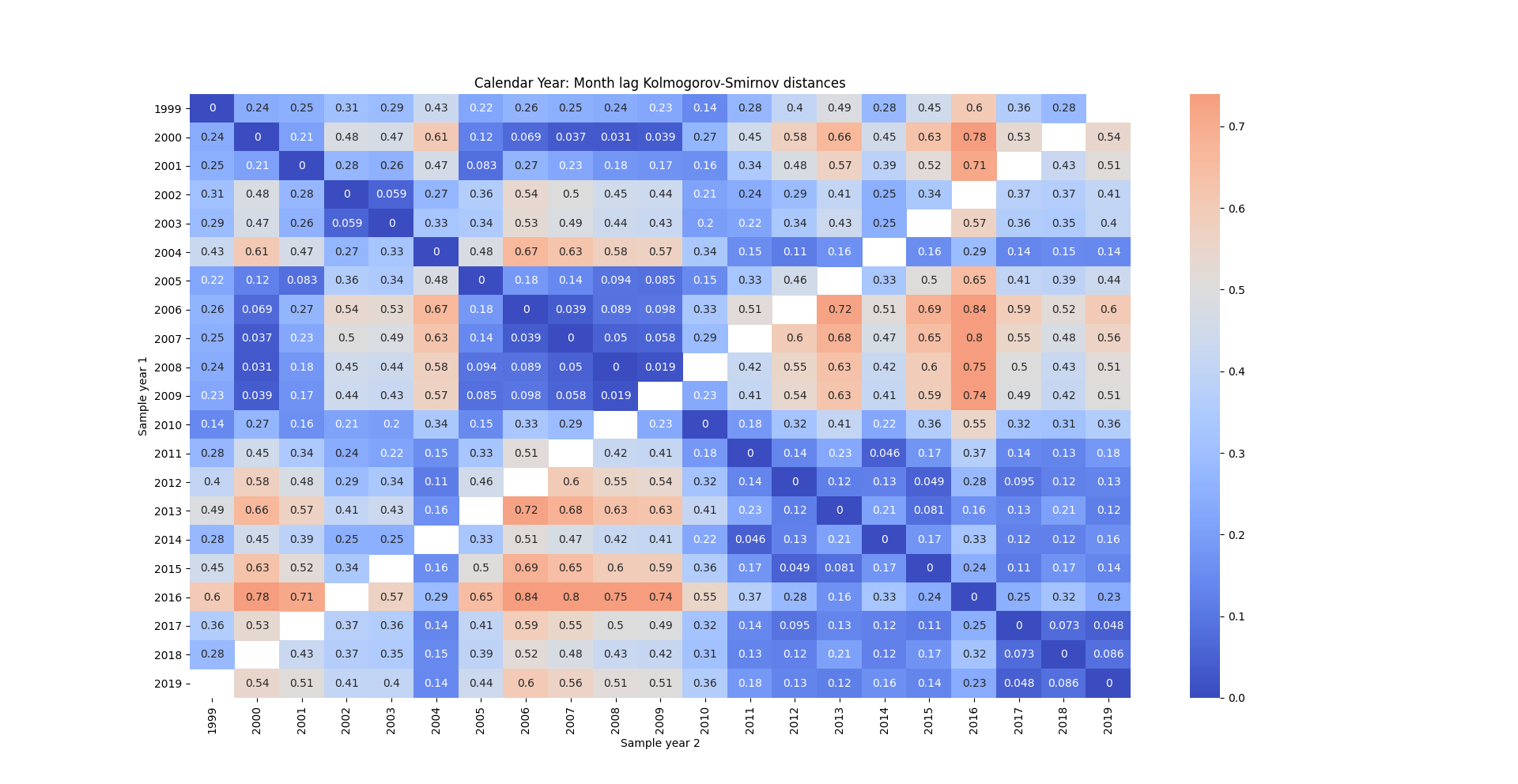}
    \caption{Kolmogorov-Smirnoff Distances in the histogram of month lags from assignment to publication.}
    \label{fig:KSCalMon}
\end{figure*}

It is important that when we view survival analysis over time in this way, we check that the averages are consistent with the calendar years. In a statistical sense we can formalise that with two sample Kolmogorov-Smirnoff tests displayed in Fig~\ref{fig:KSCalMon}. This diagram is made up of all the month lags from each calendar year in our study. Note that the number of months to publication can exceed that calendar year, but we are merely using the calendar year as a filter on the assigned date, rather than on publication.

\begin{figure}
    \centering
    \includegraphics[width=0.4\textwidth]{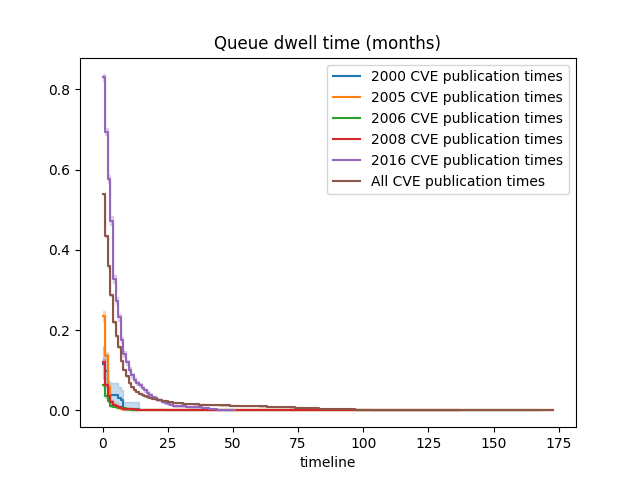}
    \caption{Kaplan-Meier Months Variances.}
    \label{fig:KMMonthVar}
\end{figure}

If we then view the years that have the furthest Kolmogorov-Smirnoff distances from each other as Kaplan-Meier curves again (See Fig~\ref{fig:KMMonthVar}, we can clearly see that a cumulative curve splits the difference of the years with the highest distance from each other. This leads us to believe that using a cumulative approach to publication time might serve us best, and keep our predictions accurate.

\begin{figure*}
    \centering
    \includegraphics[width=\textwidth]{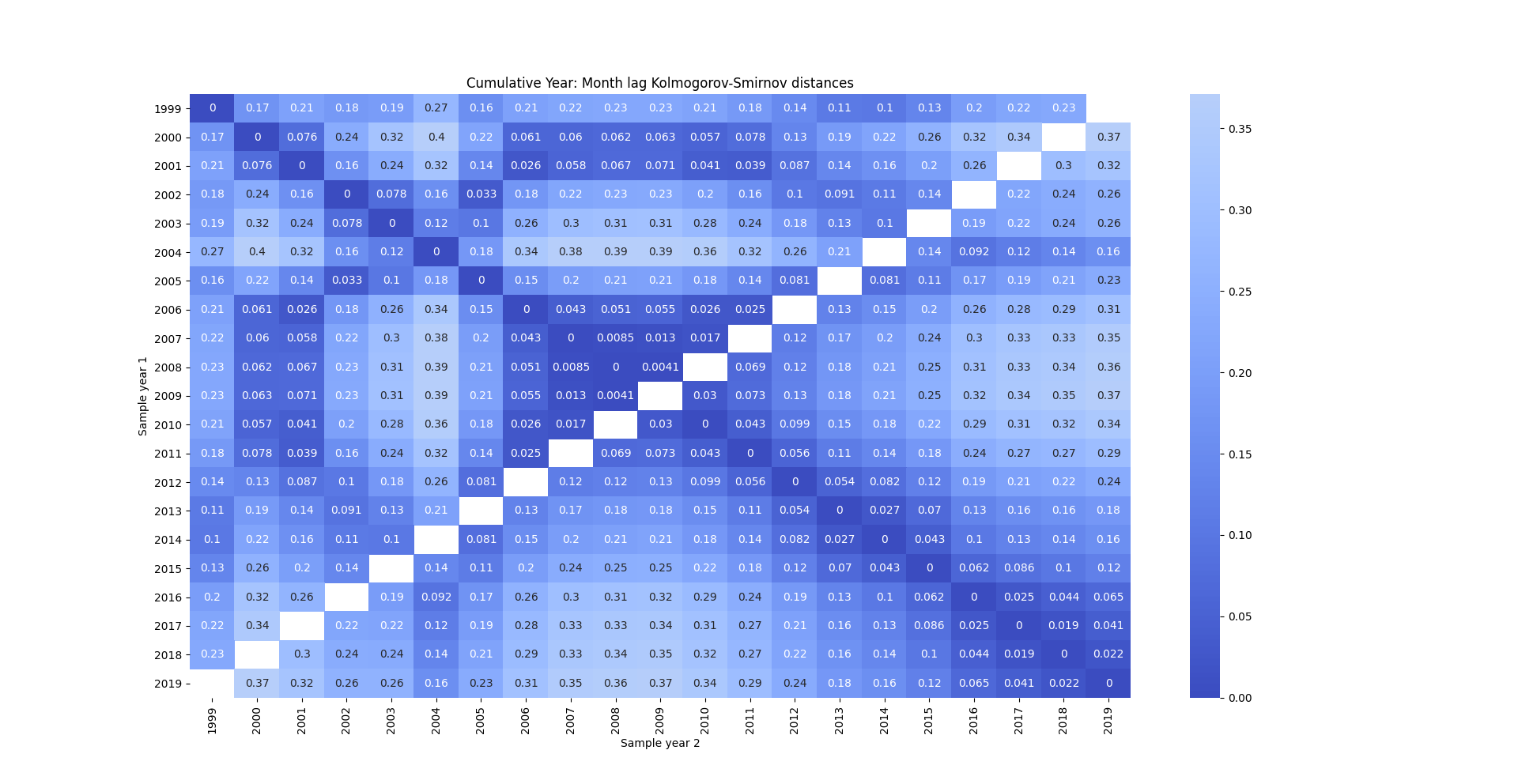}
    \caption{Kolmogorov-Smirnoff Distances in the histogram of cumulative month lags from assignment to publication..}
    \label{fig:KSCumMon}
\end{figure*}

This intuition is confirmed by viewing the Kolmogorv-Smirnoff distance graph again using the cumulatively produced record and comparing it against calendar years (See Fig~\ref{fig:KSCumMon}. 

This rather complex explanation verifies a simple result: by gathering the cumulative publication duration, and we can use them as a probabilistic estimate of publication date given assignment date. The cumulative record gives us confidence we won't be far from any individual year's publication delays, regardless of how many employees MITRE adds or how much more efficient their process becomes. CNA specific analysis could of course be conducted, but this result seems to suggest it wouldn't make a more accurate forecast. Though perhaps MITRE might like to use the same technique on each CNA to predict their workload.

While this method is easy for managing the state transitions from assigned or probable vulnerabilities to published vulnerabilities, it is quite slow when applied to all vulnerabilities in the historical record (about 7 mins on a standard laptop). Perhaps 7 minutes for a year's prediction is not unreasonable, but in practice it slowed down our retrocasting immensely. 

Initially we assumed that it was important to examine all history, because some vulnerabilities clearly come out 17 years later! However, a handful of vulnerabilities is a very minor difference in MSLogE terms, which drove us to discard larger and larger amounts of the history.

To calculate the number of vulnerabilities that are likely to be published within a year we may take a matrix, \(X\) of size \((N, M)\) where the number of years \(M=4\) since we have 3 years of history and the current year and N is the number of \textit{inferred} unpublished vulnerabilities, inferred from the highest CVE-ID seen this year. \(X\) is populated with the probability of being published in year \(m\) according to the probability vector described above such that the total number of CVEs expected this year is \(\sum{X(N, m)}\).  

This MVUE indicator operates doesn't respect calendar years, which is fine for predicting how many CVEs have been assigned this year. We have seen how knowing the number assigned can be scaled by probabilities of being published within a given timeframe. Furthermore, at the very start of the year, it may be that the few CVEs published have low ID numbers so that we underestimate the initial prediction for the CVE-YEAR, which we verified in practice; the worst MVUE predictions are in the first few weeks of January. To adapt this estimator into a predictive model two modifications are used: (1) since yearly predictions increase monotonically since the CVE-List began, we take the maximum of the current CVE-projection and the previous years; thus mitigating the issue of low scores at the start of the year. (2) in order to scale the predictions relative to the time span in question (e.g. 6 months) the MVUE is multiplied by the fraction of the year being used as the lookahead (Which is produced by examining the historical record).

Yet once again, in practical experimentation; we found we could be even more accurate \textit{and} faster, which is detailed in the next model we constructed.

\subsection{Little's Law with LookBack}

John Little's theorem, known as Little's Law, is used in queuing theory to estimate the size of a servicing queue, \(L\) from just two statistical metrics: the arrival rate into the queue, \(\lambda\), and the mean service time, \(W\) such that \(L = \lambda W\). Before a CVE is published, it is entered into the MITRE database, this can be seen as the service time. Unlike the general use of Little's law, we are not interested in how many vulnerabilities are being serviced but instead how many are going to be serviced in a given window. For this we adopt a variation of Little's Law to estimate the exit rate from the queue \(E\) using the probability that the service time is within the window of interest \(p(W \leq window)\). \(\lambda\) gives the arrival rate to the queue during the window but some items are already in the queue for the period we are predicting and we can simply count these and use this constant, \(q\) to adjust the prediction such that \(E = (q + \lambda)*p(W \leq window)\)

\(W\) is calculated from the time it takes to publish a CVE once entered into the MITRE database. 

The summary statistics for the entire NVD database (see Table~\ref{tab:lag_summaries}) shows that at the time of writing the median service time is 51 days and 16 hours but the mean is 162 days implying that the data is skewed by some large outlier values, as shown by the max of 6,632 days. Some CVEs created shortly after the CVE programme was instigated in 2001 are still being published today and these CVEs can skew both the estimation of service time and the \(q\) constant, as these CVEs are not necessarily being actively processed. For this reason we set a "lookback" limit on the Little's Law adaptation model such that the \(p(W \leq window)\) is calculated within a pre-specificed window and \(q\) only represents items added to the queue within the previous lookahead window e.g. if we are predicting how many CVEs will come out next year, we just look at how many are still left in the queue from last year and do not include the forgotten souls from the early 2000s. 

\section{Accuracy and Comparison}

This section outlines the results of our findings by applying these models. For evaluation, predictive models are compared using retrocasting for the period September 2018 to September 2020, making predictions every 10 days such that there are 72 end dates (dates to predict total CVEs by) in the test set, 24 used for validation (January to September 2018) and between 36 and 547 for training depending on the volume of historical data used.

\subsection{Predicting total CVEs}\label{framework}

To predict the total number of CVEs these experiments compare the models detailed in the previous section and also experiment with using \textbf{different historical data}: since 2005, since 2015 and since 2017 due to the large shifts in in annual figures around these years (see Fig.~\ref{fig:annual_counts}). 

for the machine learning models \textbf{different input features} are tested using features produced by the statistical models. Two of the statistical models we have developed use leading indicator data: CVE-IDs and entry/assigned dates are used by MVUE and the Little's Law with LookBack models respectively. We fed the data used by these models into the machine learning models, hypothesising that it will improve their predictions. Finally all data are combined together with count data for four different intervals: 1 week, 1 month, 1 quarter and 1 year.

Retrocasting allows us to average models performance over time historically and thus estimate how they might perform in the future. The lookahead windows tested are 1 month, 1 quarter, 6 months, and 1 year. 

\begin{figure*}
    \centering
    \includegraphics[trim={0 11cm 0 0},clip,width=0.9\textwidth]{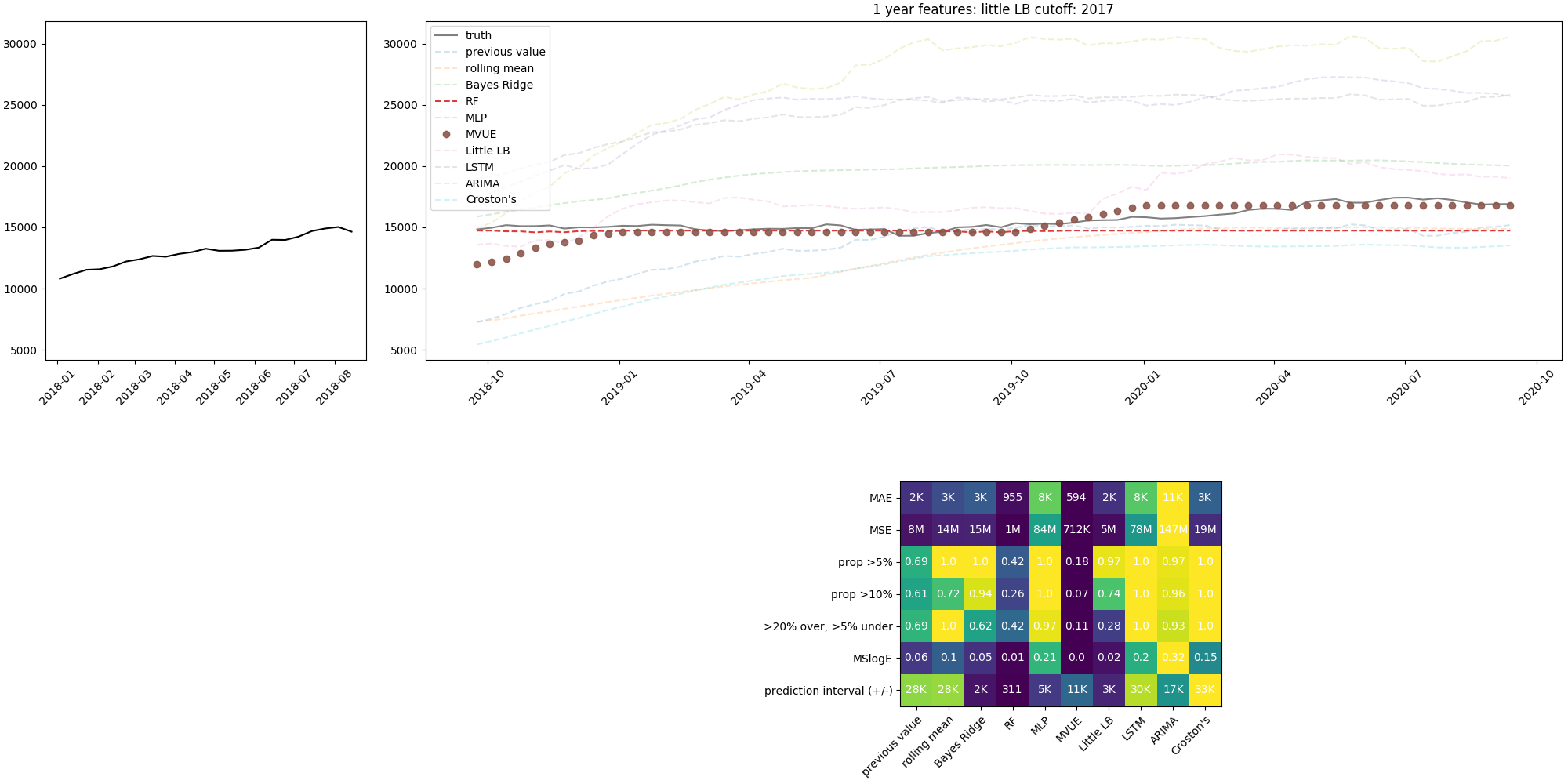}
    \caption{1 year predictions September 2018 - September 2020 using data since 2017 using Little LB features, with training data true values on the left}
    \label{fig:2017_12_months_little}
\end{figure*}

\begin{figure*}
    \centering
    \includegraphics[width=0.9\textwidth]{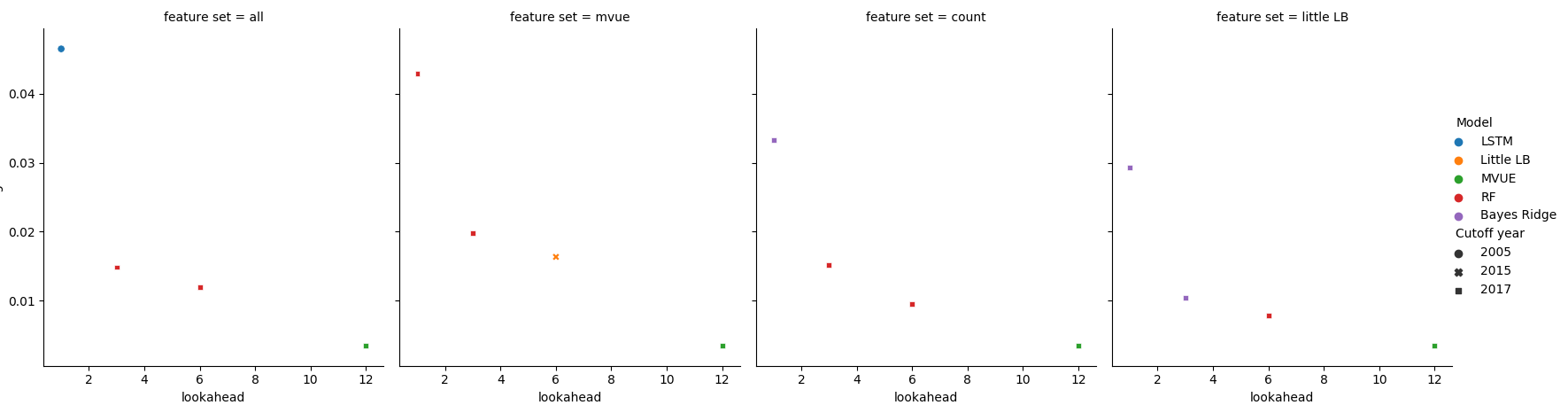}
    \caption{Best model for retrocasting by feature set}
    \label{fig:best_features}
\end{figure*}

\begin{table}
    \centering
    \begin{tabular}{lrrlr}
    \toprule
             Model &    MSlogE &  Cutoffyear & featureset &  lookahead \\
    \midrule
        Bayes Ridge &  \textbf{0.029326} &        2017 &   littleLB &          1 \\
       rolling mean &  0.046733 &        2005 &      count &          1 \\
     previous value &  0.051252 &        2005 &      count &          1 \\\hline
        Bayes Ridge &  \textbf{0.010346} &        2017 &   littleLB &          3 \\
     previous value &  0.025126 &        2017 &      count &          3 \\
       rolling mean &  0.026234 &        2017 &      count &          3 \\\hline
    Random Forest (RF) &  \textbf{0.007847} &        2017 &   littleLB &          6 \\
       rolling mean &  0.017287 &        2017 &      count &          6 \\
     previous value &  0.020160 &        2017 &      count &          6 \\\hline
        MVUE &  \textbf{0.003493} &        n/a &      n/a &         12 \\
     previous value &  0.014736 &        2015 &      count &         12 \\
       rolling mean &  0.016910 &        2015 &      count &         12 \\\hline
    \bottomrule
    \end{tabular}
    \caption{Models with lowest MSLogE predictions for lookaheads of 1, 3, 6 and 12 months compared with using the previous value or rolling mean for prediction}
    \label{tab:best_msloge}
\end{table}

Figure~\ref{fig:2017_12_months_little} shows a single retrocasting experiment to predict one year in advance using training data since 2017 and Little's Law Lookback features. To compare all retrocasting experiments, the models with the lowest MSLogE scores have been plotted by feature set (Figure~\ref{fig:best_features}) and by retrocasting cutoff year. These are compared against the previous values and rolling mean predictions in Table~\ref{tab:best_msloge}. The MSlogE is significantly reduced using the models than by using the mean or previous value. 

\begin{figure*}[!h]
    \centering
    \includegraphics[width=0.9\textwidth]{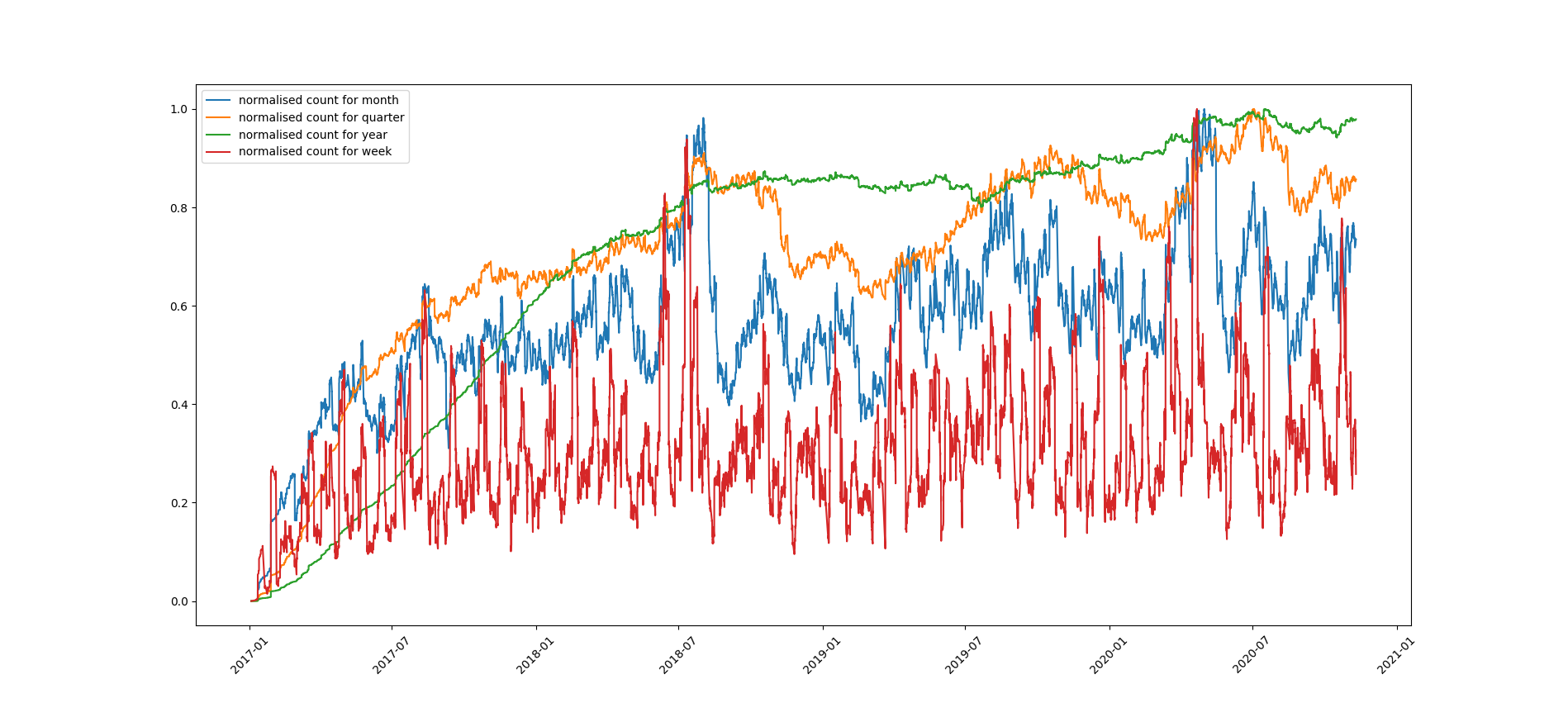}
    \caption{Weekly, monthly, quarterly and annual total vulnerabilities January 2017 to October 2020 normalised between 0 and 1 shows greater variance as for shorter time windows.}
    \label{fig:normed_variance}
\end{figure*}

In each instance, the best model has decreasing MSlogE as the lookback window increases. This is intuitive when we look at the variance in vulnerability counts over time for one-month vs one year predictions, see Figure ~\ref{fig:normed_variance} shows that by year we can see that there are exponential-looking reductions in error over time using data since 2015 and 2017 for the best models. For 2001 and 2005 we see an increase in error for 6 month predictions.

Regarding features, Figure~\ref{fig:best_features} shows that a cutoff year of 2017 for machine learning training data gives the best predictions for all lookahead windows and the Little LB output as a new data point gives the best models for 1, 3 and 6 month predictions using machine learning models despite having less data; likely because the distribution of the training and test data is more similar. 

The lowest error is for a 12 month prediction uses the MVUE model. Taking the mean for the testing period the predictions are within 3\% of the true values. For 6 months the best model is a Random Forest using Little LB features (within 7\% of true values), Naive Bayes with Little LB features is the best model for 3 months (within 8\%) and 1 month (within 12\%) predictions. Please note how fascinating and counter-intuitive it is in a forecasting sense, that we are \textbf{more accurate at longer lookahead windows}. This is a bold and confusing claim, and deserves more explanation.

%Perhaps that is an anomaly, simply a fluke of prediction interval or confidence interval size? 
Fig.~\ref{fig:normed_variance} addresses this by showing the variance in different time units. It's clear visually that the volatility reduces when you look across years instead of months or quarters. In short we are visualising the uncertainty of the \(E\) term, the exit rate, of Little's Law with Lookback. The deeper point we are trying to make is that we were surprised, and that if you merely looked at a partial auto correlation function of numbers across months, you might choose only to try and predict a 3 month lookahead as correlation is insignificant between months more than 3 months apart. That is a sort of false summit because after you cross the uncertainty of 3-9 months, the yearly predictions become useful and tractable. This is then compounded with our ability to use the MVUE/CVE numbering as another anchor to reduce our uncertainty. A third element increasing this accuracy is that Little's Law uses a longer lookback for a longer lookahead, and that too shows an increase of accuracy.  

%This is a fluke of both the MVUE and Little's Law with LookBack produced features, and a welcome one.

\subsection{Predicting subgroups of CVEs}\label{cve_types}

In practice, patch managers may want to know the forecasted total vulnerabilities for a sense of the global trends but will also be interested in the number of expected vulnerabilities for particular types of vulnerabilities such as those relating to products used by their organisation and supply chain; those vulnerabilities with a high CVSS score; those which constitute network-based attacks etc. As mentioned earlier in the paper, the vulnerabilities most likely to be exploited also represent a subcategory of interest. 

We adapted the above methodology for predicting all vulnerabilities to predict sub-categories instead. The best ML algorithm is selected using a validation set for which the selection criteria may be chosen by a patch manager. Here we select the model with the lowest MSlogE on the validation set. Not all models developed are appropriate for type forecasting - MVUE for example cannot use the CVE-IDs to predict population subsets. In these experiments we use the data-driven models with Little LB features since 2017 as these gave the best performance across all lookahead prediction windows. 

\begin{figure*}
    \centering
    \includegraphics[width=0.9\textwidth]{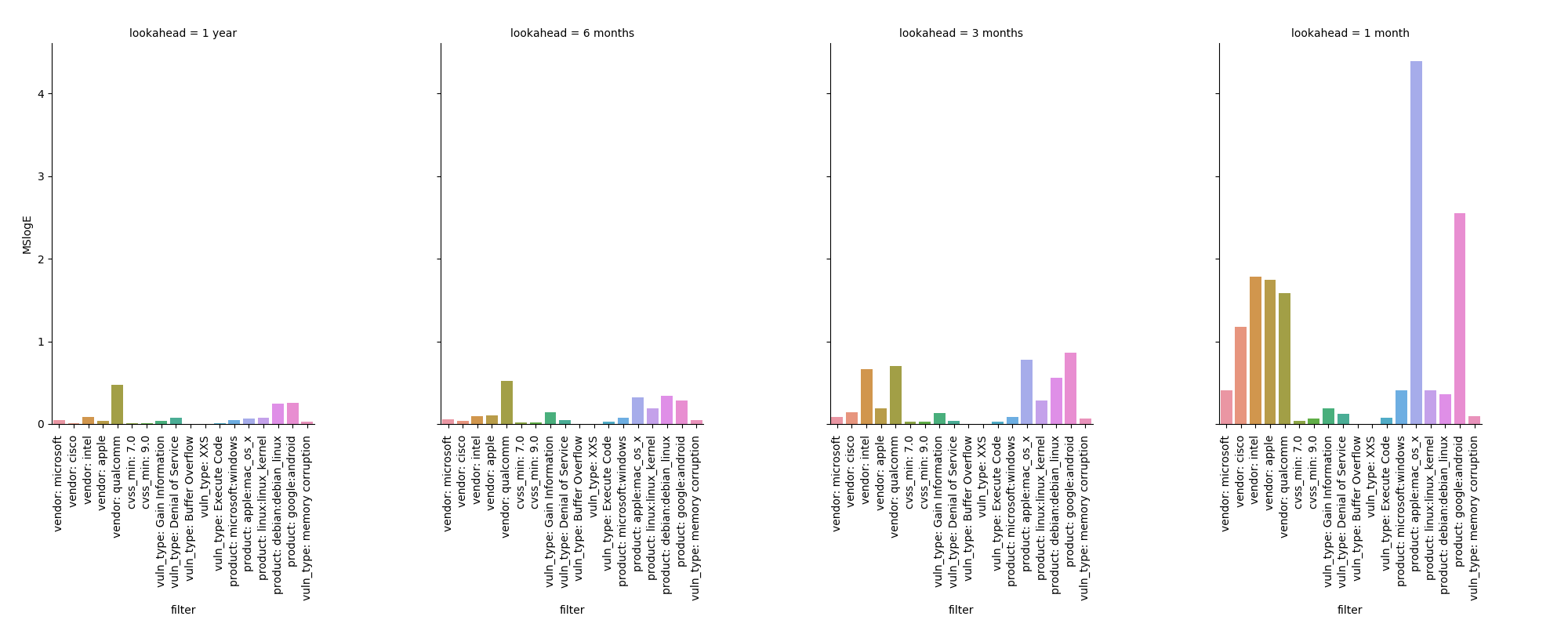}
    \caption{Forecast number of CVEs for top 5 vendors, products, vulnerability types and for CVEs with a HIGH or CRITICAL CVSS retrocasting September 2018 - September 2020}
    \label{fig:types_summary}
\end{figure*}

\begin{figure*}
    \centering
    \includegraphics[width=0.9\textwidth]{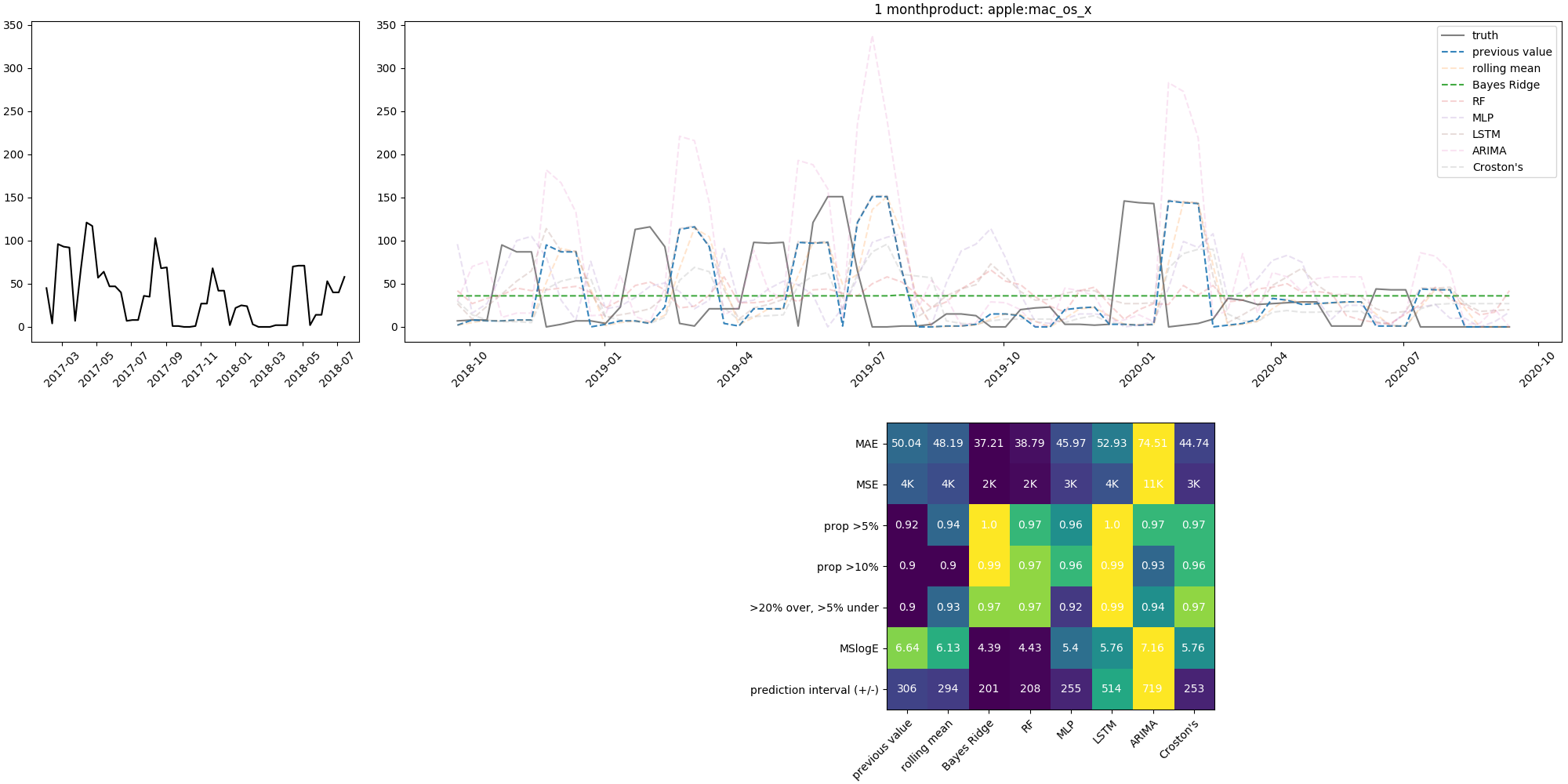}
    \caption{Mac OS X vulnerability forecasts one month ahead Sep 2018 to Sep 2020}
    \label{fig:mac}
\end{figure*}

Figure \ref{fig:types_summary} gives the MSlogE for the forecasting number of CVEs for the most common 5 products, vendors and vulnerability types seen in 2020 and for CVEs with a HIGH or CRITICAL CVSS. We can see the mean squared error is decreasing for most subtypes as the lookahead window increases. The one months prediction for Mac OS X operating system vulnerabilities stands out as having a very high MSlogE. Looking at the raw data reveals that per month the Mac OS X vulnerability count is liable to jump between 0 and 150 regularly (see Figure~\ref{fig:mac}). 

Keeping this in mind, the patch manager may wish to keep a visualisation of the variance of vulnerability numbers to hand as some products will fluctuate wildly or begin to have CVEs for the first time.

\section{Will the future prove this forecast false?}

\begin{figure*}
    \centering
    \includegraphics[width=\textwidth]{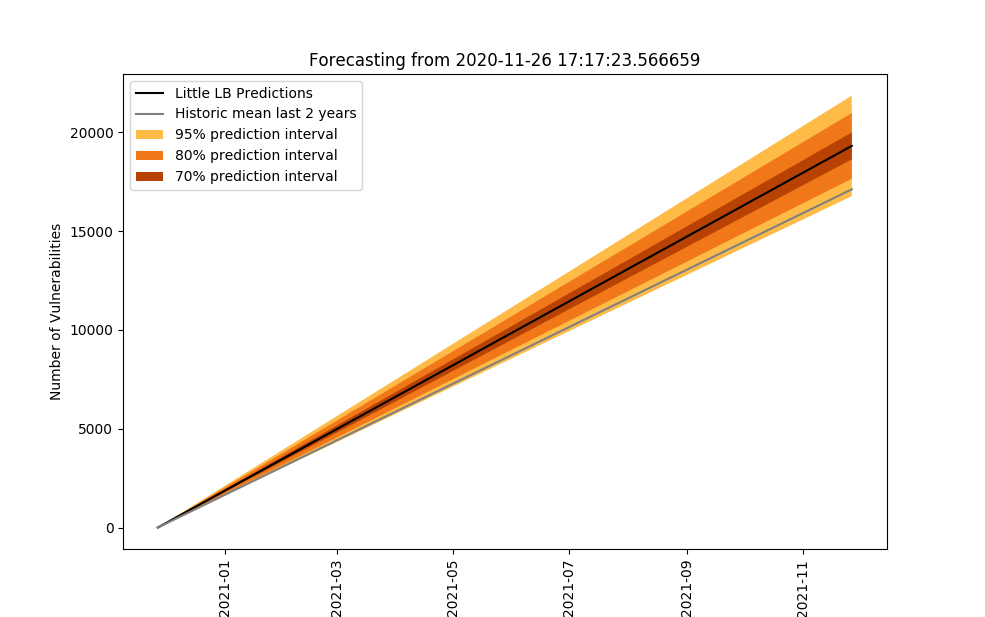}
    \caption{A forecast of the CVEs that will be published after this paper is published.}
    \label{fig:Forecast}
\end{figure*}

One sensible validation of scientific endeavour is predictive power: if a theory predicts with some accuracy things we did not know before, it validates itself. In that spirit we provide a forecast on which you can judge this paper over the next year; see Fig~\ref{fig:Forecast}. To validate it, simply search the NVD statistics \footnote{\url{https://nvd.nist.gov/vuln/search}} with a date filter from the year leading up to the Forecast date and time at the top of the image. The number of CVEs published should be within one of the prediction interval bars, and if they are not you can ask questions about how we failed to be as predictive in that lookahead. We kindly ask that you remember that the prediction accuracy we are most comfortable with is the yearly lookahead, and we hope you won't abandon us if our 3-6 month forecasts aren't as useful. If you are reading this after 2021, you will know how successful this single forecast was, and can cite us when you make better forecasts, or improve our methods.

If you're a vulnerability hunter, maybe you can invalidate the graph by finding more vulnerabilities than we expected, in which case we hope you'll cite us too. Regardless of the result, this is our effort be falsifiable and transparent.

\section{The Future of CVE Forecasting}

In 2020 (when this paper was written) the Vulnerability Management Market is estimated to be worth 12.5 Billion dollars\footnote{\url{https://www.marketsandmarkets.com/Market-Reports/security-vulnerability-management-market-204180861.html}}. It seems important that some of that money and effort be spent to give network defenders more time to patch and better specificity with respect to exploitation. To us, it seems that CVE prediction and forecasting research increases that reaction window, and thus better strategic cyber risk management decisions. Of course, one might argue that companies could be more transparent about vulnerability disclosure in general, and that would negate the research of vulnerability forecasting entirely. That is a topic for a different paper, and different researchers.

As academic research in this field makes progress both on accuracy, and extends the forecasts further into the future, it will become easier to understand the frequency of CVEs. No doubt the art will move from vulnerability forecasting to exploit forecasting. That challenge too has already been met by many\cite{jacobs2019exploit, fang2020fastembed, twitterpredict, chenTwitter}, and others even showed how long we might reasonably expect to wait to see an exploit written\cite{householder2020historical,chen}. Allodi showed us insight into the distribution of exploitation by CVE\cite{HeavyExploitation} which could be used to weight estimated exploitation before actual exploitation weights are recorded for risk purposes. Everywhere uncertainty around vulnerabilities, exploits, and exploitation is reducing, and cyber risk is maturing accordingly. We feel lucky to be a small part of that story, by predicting the volumes of CVEs.

Research then will naturally turn towards discussion of severity with respect to exploitation. When we say impact here, we do not mean in a CVSS machine specific sense. We mean 'How much will it cost me if this vulnerability is exploited in my organisation or one of my suppliers?'. This is a trickier question to answer than it initially appears, and is often bound up in the value of the assets and the business or organisation itself, more than endogenous to the statistics of vulnerabilities. It is a story of market share and computational value.

An exploit may be the mechanism of entry, but it is rarely used as a payload to gain access and then do nothing at all. These leaves two difficult to estimate problems, which essentially reduce to predictions or estimations in their own right:

\begin{enumerate}
    \item "What does the hacker want or intend after using this CVE?" (Predict what the payload will be).
    \item "How much value does any given computer or network of systems in our organisation generate?" (Predict how much we will earn or lose with/without this computer or network).
\end{enumerate}

To illustrate: some hacks don't drive a cost, and others do. If your credit card details are stolen, then my gain is your loss in a zero sum sense. However, some hacks produce non-zero sum economics. For example, if your computer is compromised and used to send spam, you may not have been using those process cycles or bandwidth. Let us be clear, this is not to excuse computer crime, but rather to illustrate that estimating losses != gains for many cyber criminal enterprises. Their utility function in an economic sense is not always antithetical to ours, though it can be.

Solving those problems is more about economics of internet enabled business models, than it is about the statistics of vulnerabilities. Even if we don't yet know what it will cost us, we are closer to understanding how many ways we have to defend ourselves now that we can construct CVE forecasts in theory, and in practice.

% \subsection{Problems our research has motivated/uncovered}

% \begin{itemize}
%     \item How do you apply a probability to products that have never had a CVE, that they might get one in the future?
%     \item Conditional probabilities and rescaling
%     \item Product/Vendor/CVSS/Type as a proportion or as a single time series?
%     \item 
% \end{itemize}

%%
%% The acknowledgments section is defined using the "acks" environment
%% (and NOT an unnumbered section). This ensures the proper
%% identification of the section in the article metadata, and the
%% consistent spelling of the heading.
%\begin{acks}
%This research and both authors were supported by Endeavr Wales, as part of an Airbus CyberLab program on AI and Cyber Defence. We would like to thank Luke Dandy for program management, encouragement, and making us explain both the vision and the science. Ivor, Mehmet, and Dilara, motivated us with smiles at the end of long days. This is what we did with our lockdown.
%\end{acks}

%%
%% The next two lines define the bibliography style to be used, and
%% the bibliography file.
\bibliographystyle{ACM-Reference-Format}
\bibliography{main}

%%
%% If your work has an appendix, this is the place to put it.
\appendix

\end{document}